\newif\ifpdflatex    
\pdflatextrue           

\documentclass{aa}
\usepackage{amsmath,esint}
\usepackage{natbib,twoopt}

\usepackage{natbib}
\usepackage{soul, CJK}
\usepackage{multirow}
\usepackage{tablefootnote}

\usepackage{bm}
\usepackage{xspace}
\usepackage{color}
\usepackage{enumitem}
\usepackage{booktabs}

\usepackage{graphicx}	
\usepackage{amsmath}	
\usepackage{amssymb}	
\usepackage{bm}		
\usepackage{url}
\usepackage{amsbsy}
\usepackage{xspace}
\usepackage{hyperref}
\usepackage{longtable}
\usepackage[colorinlistoftodos]{todonotes}
\usepackage[flushleft]{threeparttable}

\usepackage[T1]{fontenc}
\usepackage{ae,aecompl}

\def\lesssim{\mathrel{\hbox{\rlap{\hbox{\lower5pt\hbox{$\sim$}}}\hbox{$<$}}}}
\def\gtrsim{\mathrel{\hbox{\rlap{\hbox{\lower5pt\hbox{$\sim$}}}\hbox{$>$}}}}

\usepackage{upgreek}                                                  
\usepackage{xspace}
\usepackage{txfonts}
\newcommand{\um}{$\upmu$m\xspace}            
\def\apjl{ApJ}%
\def\apjs{ApJS}%
\def\aap{A\&A}%
\begin{document}
\title{R Coronae Borealis and dustless Hydrogen-deficient Carbon stars have different oxygen isotope ratios}
\titlerunning{Oxygen isotope ratios}
\authorrunning{Karambelkar et al.}








\author{V. Karambelkar
          \inst{1} \fnmsep\thanks{viraj@astro.caltech.edu}, 
          M. M. Kasliwal \inst{1},
          P. Tisserand \inst{2},
          G. C. Clayton \inst{3},
          C. L.Crawford \inst{3},
          S. G. Anand \inst{1},
          T. R. Geballe \inst{4}
          \and
          E. Montiel\inst{5}
          }

   \institute{Cahill Center for Astrophysics, California Institute of Technology, Pasadena, CA 91125, USA
         \and
            Sorbonne Universites, UPMC Univ Paris 6 et CNRS, UMR 7095, Institut d'Astrophysique de Paris, IAP, F-75014 Paris, France
        \and
            Department of Physics \& Astronomy, Louisiana State University, Baton Rouge, LA 70803, USA
        \and 
            Gemini Observatory/NSF's NOIRLab, 670 N A'ohoku Pl, Hilo HI 96720, USA
        \and
            SOFIA-USRA, NASA Ames Research Center, MS 232-12, Moffett Field, CA 94035, USA
             }



\abstract
{R Coronae Borealis (RCB) and dustless Hydrogen-deficient Carbon (dLHdC) stars are believed to be remnants of low mass white dwarf mergers. These supergiant stars have peculiar hydrogen-deficient carbon-rich chemistries and stark overabundances of $^{18}$O. RCB stars undergo dust formation episodes resulting in large-amplitude photometric variations that are not seen in dLHdC stars. Recently, the sample of known dLHdC stars in the Milky Way has more than quintupled with the discovery of 27 new dLHdC stars.}
{It has been suggested that dLHdC stars have lower $^{16}$O/$^{18}$O than RCB stars. We aim to compare the $^{16}$O/$^{18}$O ratios for a large sample of dLHdC and RCB stars to conclusively examine this claim.}
{We present medium resolution (R$\approx3000$) near-infrared spectra of 20 newly discovered dLHdC stars. We also present medium resolution (R$\approx3000-8000$) \emph{K}-band spectra for 47 RCB stars. We measure the $^{16}$O/$^{18}$O ratios of 7 dLHdC and 31 RCB stars that show $^{12}$C$^{16}$O and $^{12}$C$^{18}$O absorption bands, and present the largest sample of values of $^{16}$O/$^{18}$O for dLHdC and RCB stars to date.}
{We find that six of the seven dLHdC stars have $^{16}$O/$^{18}$O $<0.5$, while 26 of the 31 RCB stars have $^{16}$O/$^{18}$O $>1$. We also confirm that unlike RCB stars, dLHdC stars do not show strong blueshifted ($>200$ km s$^{-1}$) \ion{He}{I} 10833 $\AA$ absorption, suggesting the absence of strong, dust-driven winds around them.}
{We conclude that most dLHdC stars have lower $^{16}$O/$^{18}$O than most RCB stars. This confirms one of the first, long-suspected spectroscopic differences between RCB and dLHdC stars. Our results rule out the existing picture that RCB stars represent an evolved stage of dLHdC stars. Instead, we suggest that whether the white dwarf merger remnant is a dLHdC or RCB star depends on the mass ratios, masses and compositions of the merging white dwarfs. }


\date{Received December 14, 2021;}

\maketitle

\section{Introduction}
R Coronae Borealis (RCB) and dustless Hydrogen-deficient Carbon (dLHdC) \footnote{Historically, dLHdC stars were referred to as HdC stars. Here, we  follow the updated nomenclature of Tisserand et al. 2021.} stars are supergiants characterized by a peculiar chemical composition -- an acute deficiency of hydrogen and an overabundance of carbon \citep{Clayton1996, Asplund2000}. Together, they form the class of Hydrogen-deficient Carbon (HdC) stars. Their chemical compositions suggest that they are remnants of a He-core and a CO-core white dwarf (WD) merger \citep{Webbink84,Clayton12}, making them low mass counterparts of type Ia supernovae in the double-degenerate (DD) scenario \citep{Fryer2008}. In addition to peculiar chemical compositions, RCB stars experience rapid photometric declines (up to 8 mag in V band) that are attributed to dust formation episodes \citep{Clayton12}. dLHdC stars do not show any such declines or any significant infrared (IR) excess, suggesting that they do not undergo any dust formation \citep{Clayton12}. Why RCB stars produce dust while dLHdC stars do not, despite having no other known chemical differences is still a mystery. 

It has been relatively easier to identify RCB stars than dLHdC stars owing to their spectacular photometric variations. There are 128 RCB stars known in the Milky Way, while only 5 Galactic dLHdC stars were known for the last four decades \citep{Clayton12,Tisserand2020,Karambelkar2021}. However, the sample of Galactic dLHdC stars has more than quintupled in the last year with the discovery of 27 new dLHdC stars (Tisserand et al. 2021, submitted). Additionally, several new RCB stars have been discovered and observed spectroscopically at near-infrared (NIR) wavelengths (\citealt{Karambelkar2021}). 

NIR spectroscopic observations were key to identifying He-core and CO-core WD mergers as the progenitors of dLHdC and RCB stars. The K-band spectra of these stars show anomalously strong $^{12}$C$^{18}$O first overtone bands in addition to the $^{12}$C$^{16}$O first overtone bands \citep{Clayton2005}. The values of $^{16}$O/$^{18}$O in dLHDC and RCB stars cover a remarkably wide range (from 0.3 to 50), thus in all cases 1--3 orders of magnitude smaller than the solar value ($\sim$500). The large amount of $^{18}$O is thought to be produced by partial helium burning in a thin shell around the core of the WD merger remnant, which is convectively dredged up to the surface \citep{Clayton2007,Crawford2020}. 

It has been suggested that in addition to dust production, $^{16}$O/$^{18}$O ratios could be a second difference between RCB and dLHdC stars. \cite{Clayton2007} and \citet{Garcia-Hernandez2009} noted that dLHdC stars have $^{16}$O/$^{18}$O $<1$ and most RCB stars have $^{16}$O/$^{18}$O $>1$. However, this analysis is based on a small sample of only two dLHdC and five RCB stars \footnote{Note that the star HD 175893 was originally classified as a dLHdC star but is now known to be an RCB star based on an IR excess found by \citet{Tisserand2012}}. Both dLHdC stars in the sample have $^{16}$O/$^{18}$O $<1$, while three of the five RCB stars have $^{16}$O/$^{18}$O $>1$.  \citet{Karambelkar2021} measured the ratios for six additional RCB stars using medium resolution spectra and found that they had $^{16}$O/$^{18}$O$>1$. However, additional measurements of these ratios for dLHdC stars were not possible in the past because of the five dLHdC stars known at the time, only three had temperatures cold enough to show the CO bands \citep{Garcia-Hernandez2009,Garcia-Hernandez2010,Karambelkar2021}.  

In this paper, we present the largest sample of $^{16}$O/$^{18}$O values in dLHdC and RCB stars measured to date. We use medium resolution NIR spectra to measure $^{16}$O/$^{18}$O for 7 dLHdC and 31 RCB stars. We also present \emph{JHK}-band spectra of 21 (20 new, 1 previously known) dLHdC stars for the first time. Our results conclusively show that most dLHdC stars have lower $^{16}$O/$^{18}$O than most RCB stars. In Section \ref{sec:data}, we describe our spectroscopic observations, which include data collected over the last 15 years. In Section \ref{sec:hdc_nir_spec}, we describe the NIR spectral features of dLHdC stars. We outline our methods for measuring $^{16}$O/$^{18}$O in Section \ref{sec:measurements}. We discuss the implications of the different oxygen isotope ratios for formation scenarios of RCB and dLHdC stars in Section \ref{sec:disc}, and conclude with a summary of our results in Section \ref{sec:conclusions}. 

\section{Data}
\label{sec:data}
Our data comprise medium resolution NIR spectra of 24 dLHdC stars and 47 RCB stars. The complete log of spectroscopic observations is presented in Table \ref{tab:log}.

The spectra of 18 of the 20 newly discovered dLHdC stars were taken with the Triplespec spectrograph (\citealt{Herter2008}, R $\approx$ 3000) on the 200-inch Hale telescope at Palomar observatory on several nights between June and October 2021. Two newly discovered dLHdC star (A166 and A183) were observed with the SpeX spectrograph (R $\approx$ 2500) on the 3 m NASA Infrared Telescope Facility (IRTF, \citealt{Rayner2003}) on UT 20210623. We include a previously unpublished spectrum of the previously known dLHdC star HD 137613 taken with the UKIRT Imager Spectrometer (UIST, \citealt{Ramsay2004}, R$\approx$3100) on the 3.8 m United Kingdom Infrared Telescope (UKIRT) on UT 20050309. We also include spectra of 3 other previously known dLHdC stars obtained with the Gemini Near-infrared Spectrograph (GNIRS, \citealt{Elias2006}) on the 8.1 m Gemini-South telescope in the long-slit mode with R $\approx 5900$ in 2005, reported by \citet{Clayton2007}.

The spectra of the RCB stars were collected over the last 15 years at three telescopes with four different instruments. We include spectra of 6 RCB stars obtained with Gemini South/GNIRS in September 2005 \citep{Clayton2007}. We include spectra of 2 more RCB stars observed with the Flamingos 2 spectrograph \citep{Eikenberry2004} on the Gemini-South telescope in 2015. We observed the RCB star Z Umi with GNIRS on the Gemini-North telescope in 2011. 
In 2013, we observed 9 RCB stars with IRTF/SpeX at R $\approx$ 2500. We observed 12 more RCB stars in 2014 and 26 RCB stars in 2015 with IRTF/Spex at R $\approx$ 5000.

The Triplespec and IRTF spectra were reduced using the IDL package \texttt{spextool} \citep{Cushing2004}, and were flux calibrated and corrected for telluric absorption with standard star observations using \texttt{xtellcor} \citep{Vacca2003}. The UKIRT spectrum of HD 137613 was reduced in a standard manner using the Figaro software package for flatfielding, removal of the effects of cosmic ray hits, spectral and spatial rectification of the spectral images, wavelength calibration using an argon lamp, and ratioing by the spectrum of a standard star. The Gemini spectra were reduced similarly, but used a combination of IRAF and Figaro as described in \citet{Clayton2007}.

\begingroup
\renewcommand{\tabcolsep}{3pt}
\begin{table*}
\begin{center}
\begin{minipage}{18cm}
\caption{Log of spectroscopic observations}
\label{tab:log}
\begin{tabular}{lccc||lccc}
\hline
{Name} & {Class} & {Date} & {Tel./Inst.} & {Name} & {Class} & {Date} & {Tel./Inst.} \\ 
\hline
\hline
HD 137613                 & dLHdC & 09/03/2005      &  UKIRT/UIST      & ASAS-RCB-17               & RCB   & 08/2013,06/2015 &  IRTF/SpeX         \\
HD 182040                 & dLHdC & 09/2005         &  GS/GNIRS        & ASAS-RCB-4                & RCB   & 08/2013,06/2015 &  IRTF/SpeX         \\
HD 148839                 & dLHdC & 09/2005         &  GS/GNIRS        & MACHO-401.48170.2237      & RCB   & 08/2013,06/2015 &  IRTF/SpeX         \\
HD 173409                 & dLHdC & 09/2005         &  GS/GNIRS        & ASAS-RCB-14               & RCB   & 08/2013,08/2014 &  IRTF/SpeX         \\
A183                      & dLHdC & 23/06/2021      &  IRTF/SpeX       & ASAS-RCB-18               & RCB   & 08/2014         &  IRTF/SpeX         \\
A166                      & dLHdC & 23/06/2021      &  IRTF/SpeX       & IRAS18135.5-2419          & RCB   & 08/2014         &  IRTF/SpeX         \\
C17                       & dLHdC & 25/06/2021      &  P200/TSpec      & EROS2-CG-RCB-4            & RCB   & 08/2014         &  IRTF/SpeX         \\
A223                      & dLHdC & 25/06/2021      &  P200/TSpec      & MACHO-308.38099.66        & RCB   & 08/2014,06/2015 &  IRTF/SpeX         \\
B42                       & dLHdC & 25/06/2021      &  P200/TSpec      & WISE\_J194218.38-203247.5 & RCB   & 08/2014         &  IRTF/SpeX         \\
A226                      & dLHdC & 25/06/2021      &  P200/TSpec      & OGLE-GC-RCB-1             & RCB   & 08/2014,06/2015 &  IRTF/SpeX         \\ 
C38                       & dLHdC & 29/06/2021      &  P200/TSpec      & EROS2-CG-RCB-9            & RCB   & 08/2014         &  IRTF/SpeX         \\
C20                       & dLHdC & 29/06/2021      &  P200/TSpec      & EROS2-CG-RCB-11           & RCB   & 08/2014         &  IRTF/SpeX         \\
C105                      & dLHdC & 02/07/2021      &  P200/TSpec      & WISE\_J183649.54-113420.7 & RCB   & 08/2014         &  IRTF/SpeX         \\
F75                       & dLHdC & 02/07/2021      &  P200/TSpec      & V739 Sgr                  & RCB   & 08/2014,06/2015 &  IRTF/SpeX         \\
B565                      & dLHdC & 15/09/2021      &  P200/TSpec      & V3795 Sgr                 & RCB   & 08/2014         &  IRTF/SpeX         \\
B566                      & dLHdC & 15/09/2021      &  P200/TSpec      & ASAS-RCB-5                & RCB   & 06/2015         &  IRTF/SpeX         \\
C528                      & dLHdC & 15/09/2021      &  P200/TSpec      & ASAS-RCB-7                & RCB   & 06/2015         &  IRTF/SpeX         \\
C539                      & dLHdC & 15/09/2021      &  P200/TSpec      & ASAS-RCB-16               & RCB   & 06/2015         &  IRTF/SpeX         \\
A811                      & dLHdC & 15/09/2021      &  P200/TSpec      & ASAS-RCB-19               & RCB   & 06/2015         &  IRTF/SpeX         \\
B567                      & dLHdC & 15/09/2021      &  P200/TSpec      & ASAS-RCB-20               & RCB   & 06/2015         &  IRTF/SpeX         \\
A798                      & dLHdC & 15/10/2021      &  P200/TSpec      & EROS2-CG-RCB-3            & RCB   & 06/2015         &  IRTF/SpeX         \\
C542                      & dLHdC & 15/10/2021      &  P200/TSpec      & EROS2-CG-RCB-10           & RCB   & 06/2015         &  IRTF/SpeX         \\
B563                      & dLHdC & 15/10/2021      &  P200/TSpec      & EROS2-CG-RCB-13           & RCB   & 06/2015         &  IRTF/SpeX         \\
A814                      & dLHdC & 15/10/2021      &  P200/TSpec      & FH Sct                    & RCB   & 06/2015         &  IRTF/SpeX         \\
HD 175893                 & RCB   & 09/2005         &  GS/GNIRS        & GU Sgr                    & RCB   & 06/2015         &  IRTF/SpeX         \\     
S Aps                     & RCB   & 09/2005         &  GS/GNIRS        & RS Tel                    & RCB   & 06/2015         &  IRTF/SpeX         \\
SV Sge                    & RCB   & 09/2005         &  GS/GNIRS        & V482 Cyg                  & RCB   & 06/2015         &  IRTF/SpeX         \\
ES Aql                    & RCB   & 09/2005         &  GS/GNIRS        & V854 Cen                  & RCB   & 06/2015         &  IRTF/SpeX         \\
WX CrA                    & RCB   & 09/2005         &  GS/GNIRS        & V1783 Sgr                 & RCB   & 06/2015         &  IRTF/SpeX         \\
U Aqr                     & RCB   & 09/2005         &  GS/GNIRS        & V2552Sgr                  & RCB   & 06/2015         &  IRTF/SpeX         \\
Z Umi                     & RCB   & 03/2011         &  GN/NIRI         & VZ Sgr                    & RCB   & 06/2015         &  IRTF/SpeX         \\
ASAS-RCB-11               & RCB   & 08/2013         &  IRTF/SpeX       & WISE\_J174328.50-375029.0 & RCB   & 06/2015         &  IRTF/SpeX         \\
EROS2-CG-RCB-6            & RCB   & 08/2013         &  IRTF/SpeX       & V532 Oph                  & RCB   & 06/2015         &  IRTF/SpeX         \\
V517 Oph                  & RCB   & 08/2013         &  IRTF/SpeX       & HV 5637                   & RCB   & 11/2015         &  GS/F-2            \\
V1157 Sgr                 & RCB   & 08/2013         &  IRTF/SpeX       & EROS2-SMC-RCB-1           & RCB   & 12/2015         &  GS/F-2            \\
NSV 11154                 & RCB   & 08/2013         &  IRTF/SpeX       &                           &       &                 &                    \\         
\hline

\hline
\end{tabular}

\end{minipage}
\end{center}
\end{table*}
\endgroup

\section{NIR spectra of dLHdC stars}
\label{sec:hdc_nir_spec}
\begin{figure*}
    \centering
    \includegraphics[width=0.95\textwidth]{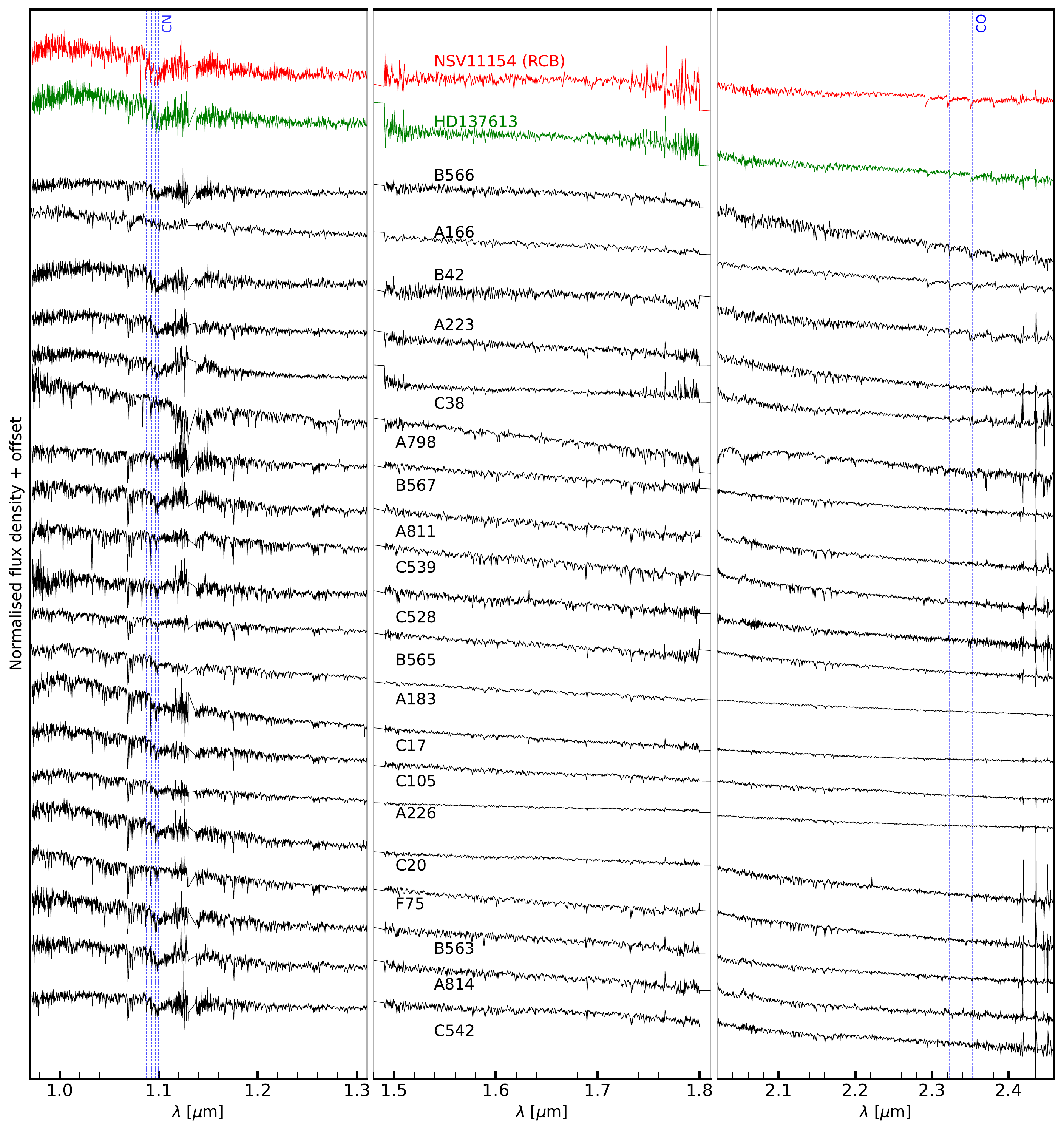}
    \caption{NIR spectra of the newly discovered dLHdC stars. NIR spectra of the RCB star NSV11154 and the previously known dLHdC star HD137613 are also shown for comparison. The rest-frame positions of the CN and $^{12}$C$^{16}$O absorption bands are indicated by dashed vertical lines.}
    \label{fig:hdc_spectra}
\end{figure*}

We present \emph{JHK-}band NIR spectra of the 20 new dLHdC stars in Figure \ref{fig:hdc_spectra}. We also include in the figure spectra of the previously known dLHdC star HD 137613 (green) and the RCB star NSV11154 (red, taken from \citealt{Karambelkar2021}). 

The spectra of the dLHdC stars closely resemble NIR spectra of RCB stars taken at maximum light. The continuum shapes resemble those of F-G type stars. Numerous absorption lines are present. We identify strong absorption features attributed to \ion{C}{I} (most prominently at 1.06883, 1.0686, 1.0688 and 1.06942 \um  in the J-band and  1.73433, 1.74533 and 1.75104 \um in the H-band) and a blend of numerous \ion{Fe}{I}, \ion{K}{I} and \ion{Si}{I} lines (see \citet{Rayner2003} for the wavelengths). H lines are absent in the NIR spectra of the dLHdC stars\footnote{See \href{https://www.gemini.edu/observing/resources/near-ir-resources/spectroscopy/hydrogen-recombination-lines}{here} for a list of NIR H lines}. Only five of the twenty new dLHdC stars show $^{12}$C$^{16}$O and $^{12}$C$^{18}$O molecular features (presumably because the others are too hot for CO to exist in detectable amounts). Of the five, all but A166 also show the $^{12}$C$^{14}$N bands at 1.0875, 1.0929, 1.0966, 1.0999 \um. Two additional stars B565 and A811 show CN but no CO bands.

\subsection{Helium 1.0833 \um triplet}
\label{sec:helium}
Our NIR spectra cover the \ion{He}{I} 1.0833 \um triplet. This feature can serve as a tracer of high velocity winds around RCB and dLHdC stars. The levels of this transition are 20 eV above the ground state, and cannot be populated by photospheric radiation of dLHdC and RCB stars. Instead, they can be collisionally excited in high velocity winds around these stars. The \ion{He}{I} feature has been observed in several RCB stars either as blueshifted absorption or with a P-Cygni profile with velocities as high as 500 km s$^{-1}$ \citep{Clayton2003,Clayton2011,Karambelkar2021}. The strength and velocity of the winds around RCB stars are greatest when the star has just emerged from a dust enshrouded minimum, and decrease with time thereafter. This suggests that the winds around RCB stars are dust-driven; i.e. the gas is dragged to high velocities by dust grains that are accelerated by radiation pressure.  As dLHdC stars do not show dust excesses in their spectral energy distributions, we do not expect to see dust-driven winds around them. \citet{Geballe2009} observed the four previously known dLHdC stars and found no evidence for strong RCB-like \ion{He}{I} absorption lines in their spectra.

\begin{figure*}
    \centering
    \includegraphics[width=\textwidth]{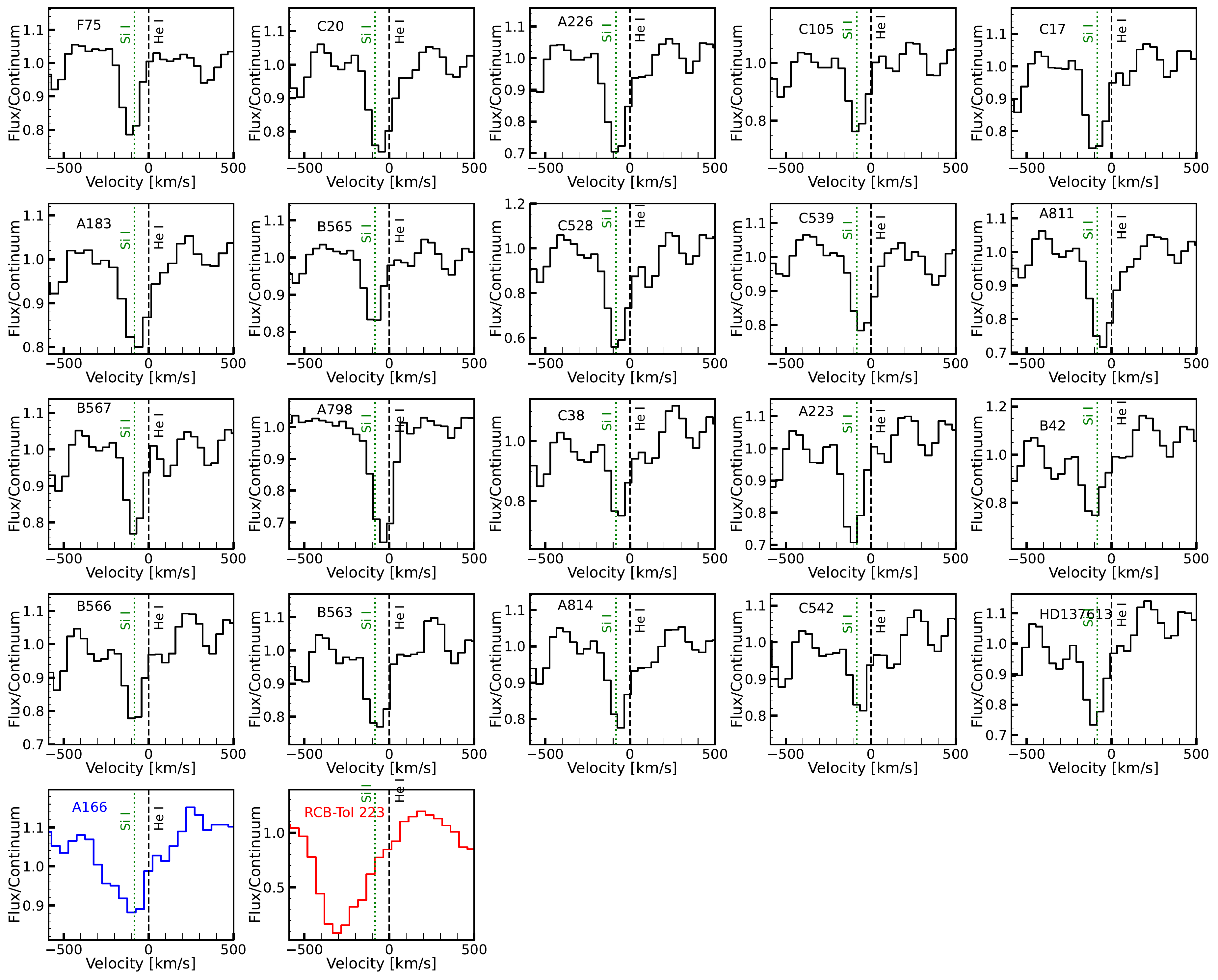}
    \caption{Zoom in of the spectral region around the \ion{He}{I} (10833 \AA) triplet. The velocity is measured with respect to 1.0833 $\mu$m. The same spectral region of \ion{He}{I} profile of an RCB star is shown in red -- RCB-ToI-223 (a.k.a. WISE J182010.96-193453.4, at a phase when it has just emerged from a dust decline). We cannot resolve the contribution from the \ion{Si}{I} line to any \ion{He}{I} features from our low resolution spectra. However, it is evident that we do not detect any strong ($>200$ km s$^{-1}$), RCB-like winds in any dLHdC star, suggesting that there is no dust-driven wind around them. This is consistent with no hot dust excess seen in the SEDs of any of these stars. A166 is the only dLHdC star that shows a possible high velocity \ion{He}{I} absorption profile (blue).}
    \label{fig:helium_profiles}
\end{figure*}

Figure \ref{fig:helium_profiles} shows a zoom-in of the NIR spectra of the new dLHdC stars around the \ion{He}{I} 1.0833 \um triplet. We do not detect RCB-like (width $>200$ km s$^{-1}$) helium absorption in the NIR spectra for any of the newly discovered dLHdC stars, except possibly A166. We cannot determine whether a lower velocity \ion{He}{I} component is present from our medium resolution spectra, as we cannot resolve possible contribution of \ion{He}{I} to the \ion{Si}{I} (1.0831 \um) line. A166 is the only dLHdC star that shows signs of an extended absorption component in addition to \ion{Si}{I} absorption. If this is indeed the \ion{He}{I} line, it would imply a wind velocity of $\approx 400$ km s$^{-1}$. Higher resolution observations are necessary to identify the sources of this absorption.  

Although we cannot completely resolve any possible low velocity \ion{He}{I} component, our observations rule out the presence of strong, RCB-like dust-driven mass loss in dLHdC stars. This is expected, as none of these stars (except A166) shows a significant IR excess (Tisserand et al. 2021, submitted). We note, however, that this does not rule out the possibility that dLHdC stars were forming dust at some point in their history ($\gtrapprox 10$ years ago). Observations of XX Cam -- an RCB star that has not entered a dust-enshrouded decline for the last six decades -- do not show any significant \ion{He}{I} features, similar to dLHdC stars \citep{Geballe2009}. The dust-driven \ion{He}{I} wind is thus only a tracer of recent (few years to decades) dust-formation.

\section{$^{16}$O/$^{18}$O : Analysis and Results}
\subsection{Analysis}
\label{sec:measurements}
In this section we constrain the $^{16}$O/$^{18}$O ratios of the five newly discovered dLHdC stars that show $^{12}$C$^{16}$O and $^{12}$C$^{18}$O bands, the previously known dLHdC stars HD 137613 and HD 182040 and 31 RCB stars that show the CO bands. Figure \ref{fig:o18band_zooms} shows the first four overtone CO bandheads of the seven dLHdC stars and six representative RCB stars. Even from visual inspection of the spectra, it is evident that in general the $^{12}$C$^{18}$O absorptions are stronger in dLHdC stars than RCB stars. Here, we describe our procedure to derive constraints on the $^{16}$O/$^{18}$O ratios from these and other spectra.

We follow the procedure described in \citet{Karambelkar2021}. We first generate synthetic spectra using a grid of hydrogen-deficient spherically symmetric MARCS (Model Atmospheres in Radiative and Convective Scheme) atmospheric models with input compositions characteristic of RCB and dLHdC stars (log $\epsilon$(H) = 7.5, C/He = 0.01, \citealt{Gustafsson1975,Gustafsson08,Bell1976,Plez2008}). We generated the synthetic spectra using the package \texttt{TURBOSPECTRUM} \citep{Alvarez1998}. We set log $g$ = 1.0 and varied the effective temperatures from 4000 to 7500 K in intervals of 250 K. We chose $^{16}$O/$^{18}$O values of 0.01, 0.05, 0.1, 0.2, 0.5, 1, 2, 5, 10, 20, 50, 500 and infinity (no $^{18}$O). We used nitrogen abundances of log $\epsilon$(N) = 7.0, 7.5, 8.0, 8.5 and 9.4. For RCB stars, the warm dust shell can contribute significantly to the K-band flux (up to 80\%, \citealt{Tisserand2013}), filling up the absorption bands. To account for this, we introduced an additional parameter f$_{\mathrm{dust}} = \frac{F_{\mathrm{shell,K}}}{F_{\mathrm{total,K}}}$, and varied it between 0 to 0.8 in steps of 0.1 for RCB stars. We do not expect this to be a significant effect for dLHdC stars as they do not have infrared dust excesses. We fit the synthetic spectra to continuum-normalised NIR spectra and visually examine each of the fits to determine the range of isotope ratios that is consistent with the observed spectra. The derived values of $^{16}$O/$^{18}$O are listed in Table \ref{tab:oratios}. 

As noted in \citet{Garcia-Hernandez2009}, it is challenging to measure the $^{16}$O/$^{18}$O ratios accurately from medium resolution spectra. The $^{12}$C$^{16}$O and $^{12}$C$^{18}$O absorption bands can be contaminated by absorption from other molecules such as $^{12}$C$^{14}$N and C$_{2}$. The absorption lines of CN and C$_{2}$ are densely packed in the K-band. It is also possible that the $^{12}$C$^{16}$O absorption bands are saturated, resulting in a smaller measured $^{16}$O/$^{18}$O ratios than the true value. There is also a degeneracy between the effects of the effective temperatures and nitrogen abundances on the depths of the CO absorption bands. For these reasons, we are not able to tightly constrain $^{16}$O/$^{18}$O, but instead provide a range of values combining the effects of the temperatures, nitrogen abundances and f$_{\rm{dust}}$. Wherever possible, we use estimates of temperatures from the literature to tighten the constraints. Despite the wide range of ratios for each star, our measurements conclusively show that most dLHdC stars have a significantly lower $^{16}$O/$^{18}$O than RCB stars.

\subsection{Results}
\subsubsection{dLHdC stars}
\label{sec:dlhdc_orats}
With the exception of A166, the depths of the $^{12}$C$^{18}$O absorption bands in the dLHdC stars are comparable to those of the $^{12}$C$^{16}$O bands (see Figure \ref{fig:o18band_zooms}). From comparisons to synthetic spectra and using temperatures from Crawford et al. 2022 (in prep), we derive $^{16}$O/$^{18}$O ratios in the ranges 0.05--0.2 for B42, C38 and B566,  0.1--0.5 for A223, HD 182040 and HD 137613. Our derived values for HD 137613 and HD 182040 are consistent with those reported by \citet{Clayton2007} from medium resolution and \citet{Garcia-Hernandez2009} from high resolution spectra. 

For B42, none of the spectral models fit the CO features at 2.349 \um and 2.352 \um. This could be because the 2.352 \um $^{12}$C$^{16}$O band is saturated. Fits to the 2.378 and 2.383 \um bands suggest $^{16}$O/$^{18}$O = 0.05--0.2, but the only models that fit these bands have a low nitrogen abundance of log $\epsilon$(N) = 7.0. The optical spectrum of B42 shows strong CN bands (see Tisserand et al. 2021), suggesting that the nitrogen abundance may not be so low. It is possible that the 2.383 \um $^{12}$C$^{16}$O band in B42 is also saturated, and the value we report is lower than the true $^{16}$O/$^{18}$O. If this is the case, our reported uncertainty on the $^{16}$O/$^{18}$O of B42 is also likely underestimated, as it does not include the systematic uncertainty associated with the saturated bands and the nitrogen abundances. Higher resolution observations are necessary to measure the true nitrogen abundance and the true $^{16}$O/$^{18}$O value of B42. 

Unlike the other dLHdC stars, A166 shows significantly weaker $^{12}$C$^{18}$O absorptions than the $^{12}$C$^{16}$O absorption; we derive $^{16}$O/$^{18}$O in the range of 20 -- 500 for this star. Tisserand et al. 2021 note that A166 is an outlier among the newly discovered dLHdC stars. In the HR diagram, it is located near two cool RCB stars and is distant from most dLHdC stars. It also has IR excesses in the WISE W3 and W4 bands. This, together with its high, RCB-like value of $^{16}$O/$^{18}$O and possible broad \ion{He}{I} absorption suggests that it is an RCB star. However, the lack of IR excesses in the NIR, WISE W1 and W2 bands indicates that it is in a phase of low dust production, similar to XX Cam (see Section 4.6 of Tisserand et al. 2021). 

\begingroup
\renewcommand{\tabcolsep}{3pt}
\begin{table*}
\begin{center}
\begin{minipage}{14cm}
\caption{Range of model parameters that best fit the observed spectra}
\label{tab:oratios}
\begin{tabular}{lccccc}
\hline
\hline
{Name} & {Class} & {Temperature} & {log($\epsilon(\mathrm{N})$)} & {f$_{\rm{dust}}$} & {$^{16}$O/$^{18}$O} \\ 
\hline
B42 $^{a}$  & dLHDC & 5000-5750  & 7.0  & 0 & $0.05-0.2$\\
C38 & dLHDC & 5000-5250  & 9.4  & 0 & $0.05-0.2$\\
B566 & dLHDC & 5500-6000  & 9.4  & 0 & $0.05-0.5$ \\
HD 137613  & dLHDC & 5000-5500 & 7.0-9.4 & 0 & $0.1-0.5$ \\ 
A223     & dLHDC & 5750 - 6000 & 9.4 & 0 & $0.1-0.5$\\
HD 182040  & dLHDC & 6000  & 8.5-9.4 & 0 & $0.1-0.5$ \\
A166      & dLHDC & 5000 - 6000  & 7.0-94 & 0 & $20-500$\\
\hline
HD 175893  & RCB & 5750 & 9.4 & 0-0.5 & $0.01-0.2$ \\
ASAS-RCB-11 & RCB & 5250 & 7.0-9.4 & 0-0.3 & $0.01 - 0.2$  \\
WX Cra   & RCB & 5250 & 9.4 & 0-0.1 & $0.05-0.5$ \\
EROS2-CG-RCB-6 & RCB & 4500-5750 & 7.0-9.4 & 0-0.8 & $0.2-5$ \\
IRAS 1813.5-2419  & RCB & 5750 & 7.0-8.5 & 0-0.5 & $0.5-10$\\
S Aps $^{a}$  & RCB & 5250 & 7.0 & 0-0.8 & $1-5$  \\
SV Sge $^{a}$  & RCB & 5250 & 7.0-7.5 & 0-0.8 & $1-5$  \\
HV 5637 & RCB & 4500-5750 & 7.0-9.4 & 0-0.8 & $1-10$ \\
ES Aql $^{a}$ & RCB & 5000-5250 & 7.0-9.4 & 0-0.3 & $2-10$  \\
Z Umi   & RCB & 5250 & 9.4 & 0.5-0.8 & $5-10$ \\
V1783 Sgr  & RCB & 5250 & 9.4 & 0.7-0.8 & $5-20$ \\
U Aqr  & RCB & 5500 & 9.4 & 0.4-0.8 & $5-20$  \\
EROS2-CG-RCB-10 & RCB & 4500-5750 &7.0-9.4 & 0-0.8 & $5-50$ \\ 
ASAS-RCB-17  & RCB & 5250 & 7.0-8.5 & 0-0.2 & $10-50$  \\
NSV 11154  & RCB & 5250 & 7.0-9.4 & 0-0.2 & $10-50$  \\
ASAS-RCB-16  & RCB & 5250 & 7.0 & 0 & $20-50$ \\
ASAS-RCB-5  & RCB & 5500 & 7.0-9.4 & 0-0.6 & $50-500$  \\
ASAS-RCB-7  & RCB & 5250 & 7.0-9.4 & 0-0.4 & $50-500$ \\
ASAS-RCB-19 & RCB & 4500-6000 & 7.0-9.4 & 0.5-0.6 & $>5$ \\
V1157 Sgr  & RCB & 5250 & 9.4 & 0-0.7 & $>10$ \\
MACHO-401.48170.2237 & RCB & 4500-6000 & 7.0-9.4 & 0-0.8 & $>20$  \\
WISE\_J174328.50-375029.0 & RCB & 5250 & 7.0-9.4 & 0-0.6 & $>20$  \\
ASAS-RCB-18 & RCB & 4750-5750 & 9.4 & 0-0.6 & $>20$ \\
OGLE-GC-RCB-1  & RCB & 5250 & 7.0-9.4 & 0-0.4 & $>50$ \\
MACHO-308.38099.66 & RCB & 4500-5500 & 7.0-9.4 & 0-0.8 & $>50$ \\
WISE\_J194218.38-203247.5 & RCB & 5500 & 9.4 & 0-0.1 & $>50$  \\
ASAS-RCB-4  & RCB & 5250 & 7.0 & 0 & $>$500  \\
EROS2-CG-RCB-13 & RCB & 4500-5500 & 7.0-9.4 & 0-0.8 & $>500$ \\
EROS2-CG-RCB-3 & RCB & 6000-6500 & 9.4 & 0-0.3& $>500$ \\
EROS2-CG-RCB-4 $^{b}$ & RCB & 4500-5500 & 7.0-9.4 & 0-0.8 & $>500$ \\ 
V517 Oph & RCB & 4500-5500 & 7.0-9.4 & 0-0.8 & $>500$  \\
\hline
\hline
\end{tabular}
\begin{tablenotes} 
\item $a$ : The $^{12}$C$^{16}$O absorption bands are possibly saturated, and the $^{16}$O/$^{18}$O is likely underestimated.
\item $b$ : EROS2-CG-RCB-4 shows $^{13}$C$^{16}$O absorption bands.
\end{tablenotes}

\end{minipage}
\end{center}
\end{table*}
\endgroup

\subsubsection{RCB stars}
\label{sec:rcb_orats}
Of the 31 RCB stars that show the CO overtone absorption bands, estimates of effective temperatures are available for 19 (Crawford et al. 2022, in prep). We are able to derive the tightest constraints on $^{16}$O/$^{18}$O for them. For most of the remaining 12, we can only derive lower limits on $^{16}$O/$^{18}$O. 

Most of the RCB stars in our sample have much weaker $^{12}$C$^{18}$O absorption bands than dLHdC stars. Five of them have $^{16}$O/$^{18}$O $\gtrapprox500$ and thus do not show $^{18}$O enrichment (compared to normal stars). A total of 26 of the 31 RCB stars have $^{16}$O/$^{18}$O $>1$. Thus, a large majority of RCB stars have higher $^{16}$O/$^{18}$O than dLHdC stars. 

\begin{figure*}[hbt]
    \centering
    \includegraphics[width=0.49\textwidth]{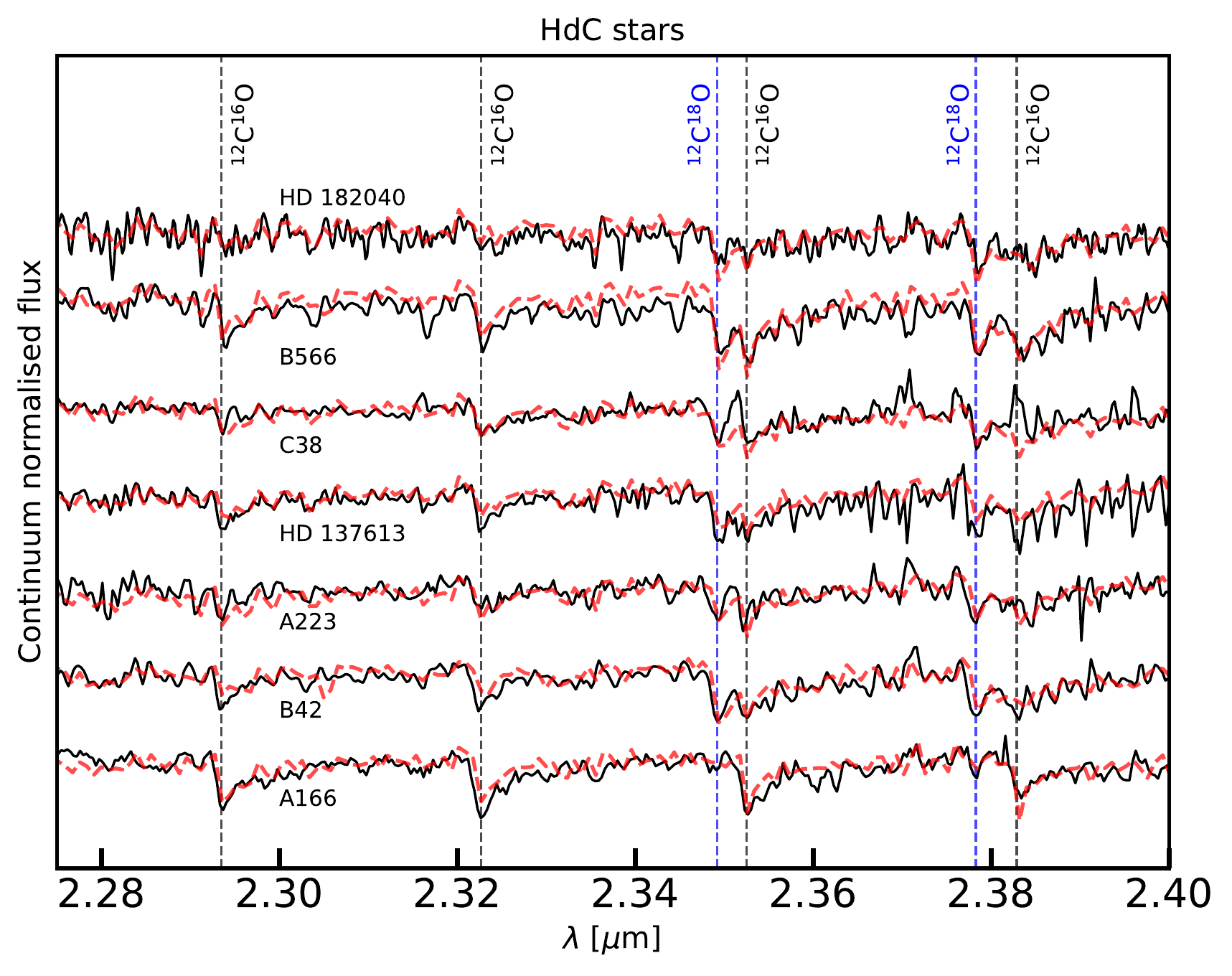}\includegraphics[width=0.49\textwidth]{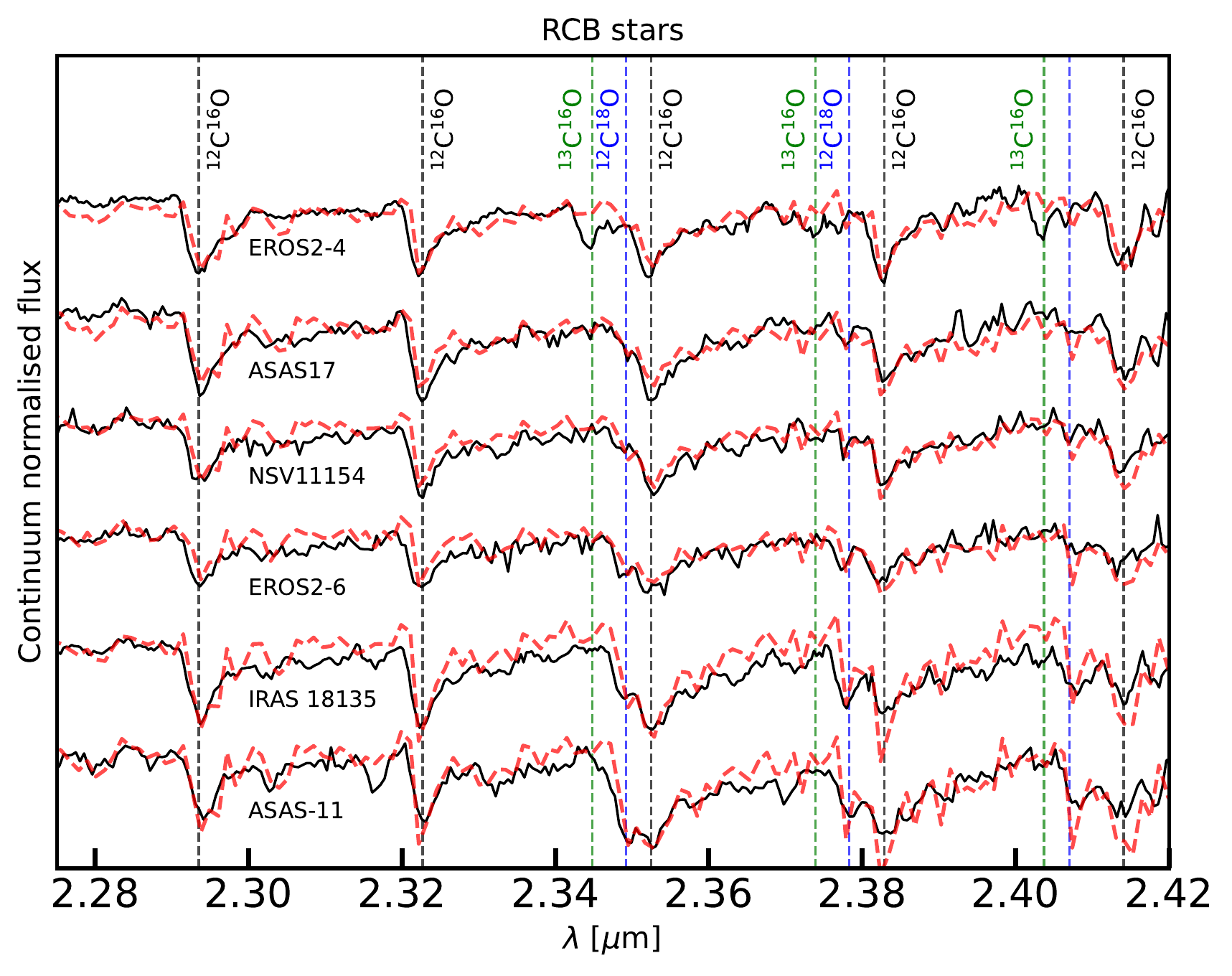}
    \caption{Zoom-in of the $^{12}$C$^{16}$O and $^{12}$C$^{18}$O absorption bands of all seven dLHdC stars (left) and six representative RCB stars (right) from our sample. All dLHdC stars except A166 show strong $^{12}$C$^{18}$O absorption, comparable in strength to the $^{12}$C$^{16}$O absorption. The RCB stars show a range of $^{12}$C$^{18}$O absorption strengths. Shown are two RCB stars where the $^{12}$C$^{18}$O absorption is strongest, two where it is of intermediate strength and two where the absorption is weak. We also plot examples of synthetic spectra that fit the observed spectra as red dashed lines. Note that the RCB star EROS2-CG-RCB4 shows $^{13}$C$^{16}$O absorption bands (right panel).}
    \label{fig:o18band_zooms}
\end{figure*}

Five RCB stars -- HD 175893, ASAS-RCB-11, IRAS 18135.5-2419, WX CrA and EROS2-CG-RCB-6 -- have prominent $^{12}$C$^{18}$O absorption features and are consistent with $^{16}$O/$^{18}$O $< 1$. We derive $^{16}$O/$^{18}$O values of 0.01--0.2, 0.01--0.2, 0.5--5, 0.05--0.5 and 0.2--5 for these stars respectively. These values are similar to those of dLHdC stars. We note that our derived $^{16}$O/$^{18}$O values for HD 175893, S Aps, SV Sge, ES Aql, U Aqr, Z Umi and WX Cra agree with previous medium resolution measurements from \citet{Clayton2007}. Our derived values also agree with high resolution measurements for all of these except S Aps, SV Sge and ES Aql, as the $^{12}$C$^{16}$O bandheads are saturated in these three \citep{Garcia-Hernandez2010}. Finally, the RCB star EROS2-CG-RCB-4 shows the $^{13}$C$^{16}$O absorption bandhead at 2.345, 2.374, 2.404 and 2.434 \um (see Figure \ref{fig:o18band_zooms}).

\begin{figure}
    \centering
    \includegraphics[width=0.5\textwidth]{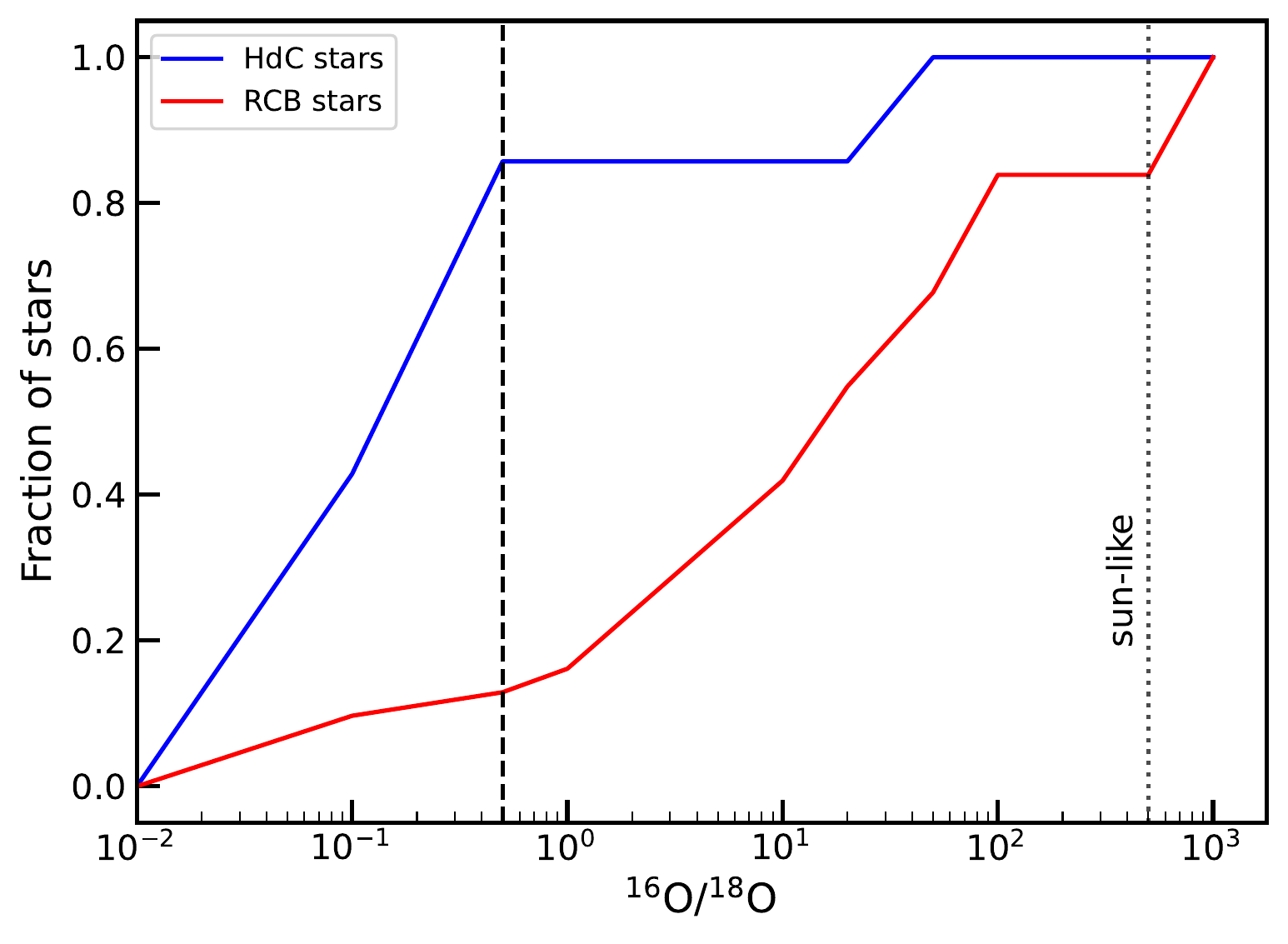}
    \caption{Cumulative distribution plot showing the fraction of dLHdC and RCB stars that have $^{16}$O/$^{18}$O below a given value. 85\% (6 out of 7) dLHdC stars have $^{16}$O/$^{18}$O $<0.5$ (black dashed line) while only $\approx15\%$ of RCB stars (5 out of 31) are consistent with such low $^{16}$O/$^{18}$O values. Most dLHdC stars have a lower $^{16}$O/$^{18}$O than most RCB stars. A166 is the only dLHdC star that has $^{16}$O/$^{18}$O $>1$. This star is an outlier amongst the newly discovered dLHdC stars, and is likely an RCB star in a low phase of dust production (see Sec. 4.1).}
    \label{fig:orats_cdf}
\end{figure}

To summarize, we find that most dLHdC stars have $^{16}$O/$^{18}$O$<1$, lower than most RCB stars. However, there is an overlap -- a small fraction of dLHdC stars have $^{16}$O/$^{18}$O $>1$ while a small fraction of RCB stars have $^{16}$O/$^{18}$O $<1$. We illustrate this in Figure \ref{fig:orats_cdf}.

\section{Discussion}
\label{sec:disc}
Our NIR spectroscopic observations have revealed that in addition to dust-formation, dLHdC and RCB stars can in most cases be distinguished based on their values of $^{16}$O/$^{18}$O. dLHdC stars in general have a lower $^{16}$O/$^{18}$O than RCB stars. It is not surprising that the oxygen isotope ratios are an important factor in the study of dLHdC and RCB stars. Anomalously low $^{16}$O/$^{18}$O in dLHdC and RCB stars was key to identifying the merger of a He-core and a CO-core white dwarf as their formation channel. $^{18}$O is synthesized by the partial helium burning reaction $^{14}$N($\alpha$,$\gamma$) F$^{18}$($\beta^{+} \nu$)$^{18}$O. This reaction is efficient at temperatures of $\approx 10^{8}$ K \citep{Clayton2007,Jeffery2011}. At higher temperatures, $^{18}$O is burnt to $^{22}$Ne. These conditions can be achieved in a thin helium burning shell around the merger remnant of a He-core and a CO-core white dwarf \citep{Clayton2007}. The $^{18}$O is convectively dredged up to the surface of the star within the first few hundred years after merger \citep{Crawford2020,Munson2021,Lauer2019}. The photospheric value of $^{16}$O/$^{18}$O in dLHdC and RCB stars is thus set within the first hundred years and remains constant for the rest of their lifetimes ($\approx 10^{4-5}$ years). Here, we explore the properties of the merging white dwarfs that set the values of $^{16}$O/$^{18}$O in the remnant supergiants. We also discuss the implications of our observations on the dLHdC-RCB connection.

\subsection{Why do dLHdC and RCB stars have different $^{16}$O/$^{18}$O?}
Studies have just begun to examine the quantities that affect $^{16}$O/$^{18}$O in white dwarf merger remnants. The effects on $^{16}$O/$^{18}$O of the helium burning shell temperature, convective extent of the supergiant envelope and the amount of hydrogen in the shell have been explored \citep{Crawford2020,Munson2021}. Their values in turn depend on the properties of the merging He-core and CO-core white dwarfs, such as their masses (M$_{\rm{He}}$ and M$_{\rm{CO}}$), mass ratios and compositions. The different values of $^{16}$O/$^{18}$O in dLHdC and RCB stars thus suggest a correlation between the progenitor white dwarfs' properties and dLHdC/RCB formation, which we discuss here. Note however that there are several additional factors that could contribute to the oxygen isotope ratios and remain to be modeled. A big unmodeled source of uncertainty is the effect of the initial composition of the progenitor white dwarfs and post-merger remnant on $^{16}$O/$^{18}$O. The correlations that we discuss below are based on current understanding and are hence preliminary. Nevertheless, they demonstrate that different properties of merging white dwarfs can result in different values of $^{16}$O/$^{18}$O, possibly providing a natural explanation for dLHdC versus RCB formation. Additional modeling will help establish robust relations between progenitor properties and dLHdC-RCB stars.

\citet{Crawford2020} explored the effect of the temperature of the helium burning shell (T$_{\rm{He}}$) on $^{16}$O/$^{18}$O. In particular, $^{16}$O/$^{18}$O drops below unity for shell temperatures in the range T$_{\rm{He}}$ $\approx 2.5-3.5 \times 10^{8}$ K. Thus, dLHdC stars could be associated with WD merger remnants that have T$_{\rm{He}}$ in this range.  For shell temperatures outside this range, $^{16}$O/$^{18}$O increases to RCB-like values, suggesting that RCB stars either have T$_{\rm{He}} < 2.5\times10^{8}$ K or T$_{\rm{He}} > 3.5\times10^{8}$ K. 

From simulations of white dwarf mergers, \citet{Staff2012} found that T$_{\rm{He}}$ is inversely correlated with the mass ratio ($q =$ M$_{\rm{He}}$/M$_{\rm{CO}}$). They derive values of T$_{\rm{He}} = 3$, 2.5 and $1.5 \times10^{8}$ K for $q=0.5$, 0.6 and 0.7 respectively. The helium shell temperatures of dLHdC stars are consistent with mass ratios $0.5<q<0.6$. RCB stars are consistent either with $q>0.6$ or $q<0.5$ depending on whether they have T$_{\rm{He}} < 2.5\times 10^{8}$ K or T$_{\rm{He}}>3.5\times10^{8}$ K, respectively. However the \citet{Staff2012} analysis does not include an important factor -- additional energy contributions from nucleosynthesis in the shell. This additional energy will increase the shell temperature, and thus the above estimates of T$_{\rm{He}}$ are likely lower limits, and the $q$ ranges are unrealistic. No studies of temperature increase due to nucleosynthesis exist in literature.

To first order, we can estimate the temperature increase timescale assuming triple-$\alpha$ burning is the dominant nucleosynthetic energy source \citep{Hansen2004}. Assuming the shell parameters from \citet{Staff2012} and a 100\% energy to temperature conversion, we find that the temperature of a $q=0.7$ merger with initial T$_{\rm{He,init}} = 1.5\times10^{8}$ K doubles within the first few years after merger. This time reduces to a few days for $q=0.6$ (T$_{\rm{He,init}}=2\times10^8$ K) and a few minutes for $q=0.5$ (T$_{\rm{He,init}}=3\times10^8$ K). The shell-burning temperatures for $q<0.6$ mergers can thus exceed $4\times10^{8}$ K. RCB stars would then be consistent with low mass-ratio ($q<0.6$) mergers and dLHdC stars would be consistent with higher mass-ratio mergers ($q>0.7$) \footnote{Note that for mergers where T$_{\rm{He}}>4.5\times10^{8}$ K, other elemental abundances predicted by models do not agree with the observed RCB abundances \citep{Crawford2020}. This suggests that there is a lower limit on the mass-ratios of white dwarf binaries that can produce RCB/dLHdC stars.}. More detailed studies of nucleosynthetic energy production are required to determine the exact ranges of \emph{q} that might form RCB and dLHdC stars 

T$_{\rm{He}}$ also depends on the total mass of the white dwarf merger \citep{Staff2018}; however, the exact dependence has not been studied extensively. The \citet{Staff2012} estimates mentioned above assumed M$_{\rm{tot}} \approx 0.9$ M$_{\odot}$. Without including nucleosynthetic energy, \citet{Staff2018} found that a M$_{\rm{tot}} = 0.7$ M$_{\odot}$, $q=0.5$ merger has T$_{\rm{He}}<2\times10^{8}$ K -- lower than the estimate of $3\times10^{8}$ K for the 0.9 M$_{\odot}$, $q=0.5$ merger. This suggests that lower mass mergers have lower T$_{\rm{He}}$ than higher mass mergers. Adding nucleosynthetic energy can increase T$_{\rm{He}}$ for the 0.7 M$_{\odot}$ merger to the dLHdC range ($2.5-3.5\times10^8$ K) and the 0.9 M$_{\odot}$ merger outside this range. dLHdC stars would then be consistent with arising from lower mass mergers than RCB stars. This would be consistent with the observation in Tisserand et al. (2021) that the population of dLHdC stars has lower luminosities than RCB stars, which they interpret as a consequence of dLHdC stars originating from lower mass mergers than RCB stars. 

In addition to T$_{\rm{He}}$, $^{16}$O/$^{18}$O can also depend on the extent of the convective envelope and the mass of hydrogen in the helium burning shell \citep{Zhang14,Munson2021}. Munson et al. explored the effect of convective overshoot factor (\emph{f}) on $^{16}$O/$^{18}$O. They found that for $f<0.07$, $^{16}$O/$^{18}$O $\approx 1$, but increases rapidly for $f>0.07$. Large values of \emph{f} correspond to cases where the convective envelope extends all the way to the CO core and dredges up additional $^{16}$O from the core into the envelope. The high values of $^{16}$O/$^{18}$O in RCB stars could be partly explained due to such a deep convective dredge up. That will also reduce C/O in RCB stars compared to dLHdC stars. The initial conditions of a merging system that would give rise to a deep convective zone remain to be explored. \citet{Munson2021} also showed that $^{16}$O/$^{18}$O increases with increasing mass of hydrogen in the helium-burning shell. The presence of hydrogen leads to a decrease of $^{18}$O and increase in $^{16}$O due to proton capture reactions. The source of this hydrogen is a thin hydrogen envelope around the progenitor He white dwarf \citep{Zhang14}. The mass of the hydrogen envelope is inversely correlated with the mass of the He white dwarf \citep{Staff2012,Driebe1998}. However, a significant fraction of this hydrogen is expected to be burned during the merger. How much of the envelope hydrogen actually reaches the helium shell is not known.

In conclusion, the $^{16}$O/$^{18}$O values in white dwarf merger remnants are affected by the properties of the merging white dwarfs such as their total mass, mass ratios and individual compositions. In this context, the different values of $^{16}$O/$^{18}$O in dLHdC and RCB stars indicate that they are formed from distinct populations of progenitor white dwarf binary mergers.

\subsection{Is there an evolutionary link between dLHdC and RCB stars?}
\citet{Garcia-Hernandez2010} proposed an evolutionary link between dLHdC and RCB stars. They note that depending on the masses and compositions of the merging white dwarfs, the merger remnant will have different initial temperatures on the supergiant track. They associate dLHdC stars with the cooler parts of the supergiant track and RCB stars with hotter temperatures. In their model, depending on merger conditions the merger may first form a cold dLHdC star and eventually evolve to an RCB star, or directly form an RCB star. 

Our new observations show that this picture is incorrect. First, several newly discovered dLHdC stars have higher photospheric temperatures than many RCB stars (Tisserand et al. 2021, submitted). Second, the spectra presented and analyzed here demonstrate that dLHdC stars in general have different values of $^{16}$O/$^{18}$O than RCB stars (Section \ref{sec:measurements}). As the surface abundances of dLHdC and RCB stars remain constant throughout their lifetimes \citep{Crawford2020}, it is not possible for one class to evolve into another.

Instead, we propose that whether a white dwarf merger forms a dLHdC or an RCB depends solely on the properties of the merging white dwarfs. The merger forms a dLHdC star (no significant dust formation, low $^{16}$O/$^{18}$O) or an RCB star (dust formation, high $^{16}$O/$^{18}$O) based on the mass ratios, masses and compositions of the white dwarfs. In this picture, it is still a mystery why RCB stars form large amounts of dust while dLHdC stars do not. It seems reasonable to expect that only those white dwarf mergers that are more massive than a certain threshold or have particular ranges of chemical compositions form remnants that can undergo dust formation. Future studies that more precisely identify the differences between dLHdC and RCB star progenitors could provide an answer to this question. 

\subsection{The overlap between dLHdC and RCB stars}
An intriguing observation is the five RCB stars (classified based on their observed IR excesses and brightness declines) that show low, dLHdC-like $^{16}$O/$^{18}$O. These ``overlap" stars may be a consequence of there being several different factors that contribute to $^{16}$O/$^{18}$O. For example, they could arise from typical RCB-like mass-ratio mergers but with a smaller total mass, smaller convective envelope or a smaller hydrogen content than typical. Another interesting possibility is that some dLHdC stars undergo a short-lived, dust forming phase. During this phase, these stars would be observed as dust-forming RCB-like stars with low $^{16}$O/$^{18}$O - similar to the five RCB stars mentioned above. This scenario does not contradict the observation that dLHdC stars do not show significant excesses in the mid-IR WISE bands. Tisserand et al. 2021 show that the emission from a circumstellar dust shell ejected from the surface of a star shifts from the mid-IR to longer wavelengths within a few decades. Thus, even if some dLHdC stars experienced dust formation $\sim10$ years before their WISE observations, they would not show any excesses in the WISE bands. The dust around them would have cooled to $<50$ K (see \citealt{Montiel2018}) and would be observable only at very long wavelengths. Only one dLHdC star has been observed at far-infrared (FIR) wavelengths to date -- HD 173409 which did not show any FIR excess \citep{Montiel2018}. FIR and sub-mm observations of the newly identified dLHdC stars will help confirm or rule out the presence of a cold dust shell around them. 


Finally, we note that four newly discovered dLHdC stars (A166, C526, F75 and F152) show definite signs of dust formation from their mid-IR colors and light curves (see Sec. 4.5 of Tisserand et al. 2021). Of these, A166 is an outlier within the dLHdC population (see Sec. \ref{sec:dlhdc_orats}), has RCB-like $^{16}$O/$^{18}$O and is likely an RCB star in a low dust-production phase. The three other stars are not outliers, but are too hot to sustain $^{12}$C$^{16}$O and $^{12}$C$^{18}$O in their photospheres so their $^{16}$O/$^{18}$O cannot be constrained. Sub-mm searches for the colder circumstellar medium of these stars can determine whether they have dLHdC-like or RCB-like $^{16}$O/$^{18}$O values. Additional studies of such ``overlap" stars are necessary to understand the dust-formation connection between dLHdC and RCB stars.

\section{Summary and way forward}
\label{sec:conclusions}
$^{18}$O has been key to understanding the origins of the enigmatic dLHdC and RCB stars. The anomalously large $^{18}$O abundance in them can be explained by invoking a He-core and a CO-core white dwarf merger model for their formation. Although there were indications that RCB and dLHdC stars have different values of $^{16}$O/$^{18}$O , the limited sample of dLHdC stars prevented a quantitative comparison. In this paper, we have utilized the revolutionary discovery of 27 new Galactic dLHdC stars to revisit this question.

We analyzed NIR spectra of 24 dLHdC stars together with unpublished spectra of 47 RCB stars. We derived $^{16}$O/$^{18}$O ratios for 7 dLHdC and 31 RCB stars whose spectra contain the $^{12}$C$^{16}$O and/or $^{12}$C$^{18}$O bands.  We find that six of the seven dLHdC stars have $^{16}$O/$^{18}$O $<0.5$, while 26 of the 31 RCB stars have $^{16}$O/$^{18}$O $>1$. This conclusively shows that most dLHdC stars have lower $^{16}$O/$^{18}$O than RCB stars. This is the only known chemical difference between these two classes of HdC stars. It remains to be seen if the lower $^{16}$O/$^{18}$O is related to the lack of dust formation in dLHdC stars.


The different oxygen isotope ratios suggest that there is no evolutionary link between the class of dLHdC and RCB stars. Instead, this observation is consistent with the picture that dLHdC and RCB stars are formed from merging white dwarfs with distinct masses, mass ratios and compositions. Further theoretical studies are required to accurately determine the properties of white dwarfs that merge to form dLHdC versus RCB stars. A small number of RCB stars have uncharacteristically low, dLHdC-like $^{16}$O/$^{18}$O values. This could be a consequence of multiple white dwarf properties that can affect the value of $^{16}$O/$^{18}$O in the merger product, or can be explained by a short-lived dust formation phase in dLHdC stars. Theoretical models will test the former scenario, while FIR and sub-mm observations will confirm or rule out the latter. Further investigations of these ``overlap" stars will shed light on the dLHdC-RCB dust formation mystery.

Future higher resolution NIR spectroscopy of the newly discovered dLHdC stars with CO bands will allow accurate determinations of their $^{16}$O/$^{18}$O ratios. High resolution optical spectroscopy should allow the determination of accurate fluorine abundances in them. As fluorine is also a signature of a white dwarf merger \citep{Pandey2007}, these observations will help determine the properties of progenitors of the hot dLHdC stars that do not show CO overtone bands. High resolution spectroscopy also will potentially identify additional chemical differences between dLHdC and RCB stars and shed further light on their progenitor white dwarf populations. 

\begin{acknowledgements}
We thank Bradley Munson for useful comments and discussions. PT acknowledges financial support from "Programme National de Physique Stellaire" (PNPS) of CNRS/INSU, France. MMK acknowledges the Heising-Simons foundation for support via a Scialog fellowship of the Research Corporation. MMK acknowledges generous support from the David and Lucille Packard Foundation. SA acknowledges support from the GROWTH PIRE grant 1545949. GC and CC are grateful for support from National Science Foundation Award 1814967. This research is based in part on observations for programs GN-2005B-Q-20 and GN-2011A-Q-112 obtained at the international Gemini Observatory, a program of NSF’s NOIRLab, which is managed by the Association of Universities for Research in Astronomy (AURA) under a cooperative agreement with the National Science Foundation. on behalf of the Gemini Observatory partnership: the National Science Foundation (United States), National Research Council (Canada), Agencia Nacional de Investigaci\'{o}n y Desarrollo (Chile), Ministerio de Ciencia, Tecnolog\'{i}a e Innovaci\'{o}n (Argentina), Minist\'{e}rio da Ci\^{e}ncia, Tecnologia, Inova\c{c}\~{o}es e Comunica\c{c}\~{o}es (Brazil), and Korea Astronomy and Space Science Institute (Republic of Korea).
\end{acknowledgements}

\bibliography{myreferences}
\bibliographystyle{apj}

\label{lastpage}
\end{document}